%Paper: hep-ph/9310322
%From: "Neil Turok" <neil@puhep1.Princeton.EDU>
%Date: Wed, 20 Oct 93 12:55:06 EDT
%Date (revised): Wed, 20 Oct 93 13:12:27 EDT
%Date (revised): Fri, 22 Oct 93 10:24:24 EDT

% Plain Tex - Just tex the file and reply b to the question.
% Colour Postscript  Figures 1 and Figures 2a-e available
% by anonymous ftp from astro.princeton.edu, in directory
% coulson/Pi
%

%%%%%%%%%%%%%%%%%%  tex macros for harvard preprints, cm version %%%%%%%

%                      P. Ginsparg, P. Nelson
%                      last updated 4/88
%                      nonharvard users say "b" in response
%                      to query, and only plain.tex is needed
%
\newbox\leftpage \newdimen\fullhsize \newdimen\hstitle \newdimen\hsbody
\tolerance=1000\hfuzz=2pt
\def\bigans{b }
\message{ big or little (b/l)? }\read-1 to\answ
\ifx\answ\bigans\message{(This will come out unreduced.}
%\magnification=1200\baselineskip=16pt plus 2pt minus 1pt
\magnification=1200\baselineskip=18pt plus 2pt minus 1pt
%\magnification=1200\baselineskip=32pt plus 2pt minus 1pt
\hsbody=\hsize \hstitle=\hsize %take default values for unreduced format
\else\def\apans{l }\message{ lyman or hepl (l/h) (lowercase!) ? }
\read-1 to \apansw\message{(This will be reduced.}
\let\lr=L
\magnification=1000\baselineskip=16pt plus 2pt minus 1pt
\voffset=-.31truein\vsize=7truein
\hstitle=8truein\hsbody=4.75truein\fullhsize=10truein\hsize=\hsbody
%	Send local landscape command to laserprinter
\ifx\apansw\apans\special{ps: landscape}\hoffset=-.59truein% apple lw
  \else\hoffset=.05truein\fi% qms laserprinter
\output={\ifnum\pageno=0 %%% This is the HUTP version
  \shipout\vbox{\hbox to \fullhsize{\hfill\pagebody\hfill}}\advancepageno
  \else
  \almostshipout{\leftline{\vbox{\pagebody\makefootline}}}\advancepageno
  \fi}
\def\almostshipout#1{\if L\lr \count1=1
      \global\setbox\leftpage=#1 \global\let\lr=R
  \else \count1=2
    \shipout\vbox{\ifx\apansw\apans\special{ps: landscape}\fi %satisfies dvips
      \hbox to\fullhsize{\box\leftpage\hfil#1}}  \global\let\lr=L\fi}
\fi
%
%---------------------------------------------------------------------
\catcode`\@=11 % This allows us to modify PLAIN macros.
\newcount\yearltd\yearltd=\year\advance\yearltd by -1900

\def\Title#1#2{\nopagenumbers\abstractfont\hsize=\hstitle\rightline{#1}%
\vskip 1in\centerline{\titlefont #2}\abstractfont\vskip .5in\pageno=0}
\def\Date#1{\vfill\leftline{#1}\tenpoint\supereject\global\hsize=\hsbody%
\footline={\hss\tenrm\folio\hss}}% restores pagenumbers
\def\draftmode{\def\draftdate{{\rm preliminary draft:
\number\month/\number\day/\number\yearltd\ \ \hourmin}}%
\headline={\hfil\draftdate}\writelabels\baselineskip=20pt plus 2pt minus 2pt
{\count255=\time\divide\count255 by 60 \xdef\hourmin{\number\count255}
	\multiply\count255 by-60\advance\count255 by\time
   \xdef\hourmin{\hourmin:\ifnum\count255<10 0\fi\the\count255}}}
%use this one instead of \Date on the preliminary draft
%it puts the current date on each page in big mode

%use \nolabels to get rid of eqn and ref labels in draft mode
\def\nolabels{\def\eqnlabel##1{}\def\eqlabel##1{}\def\reflabel##1{}}
\def\writelabels{\def\eqnlabel##1{%
{\escapechar=` \hfill\rlap{\hskip.09in\string##1}}}%
\def\eqlabel##1{{\escapechar=` \rlap{\hskip.09in\string##1}}}%
\def\reflabel##1{\noexpand\llap{\string\string\string##1\hskip.31in}}}
\nolabels
%
% tagged sec numbers
\global\newcount\secno \global\secno=0
\global\newcount\meqno \global\meqno=1
\def\newsec#1{\global\advance\secno by1
\xdef\secsym{\the\secno.}\global\meqno=1
%\ifx\answ\bigans \vfill\eject\else
\bigbreak\bigskip%\fi% (combination \goodbreak\bigskip\bigskip)
\noindent{\bf\the\secno. #1}\par\nobreak\medskip\nobreak}
\xdef\secsym{}
\def\appendix#1#2{\global\meqno=1\xdef\secsym{\hbox{#1.}}\bigbreak\bigskip
\noindent{\bf Appendix #1. #2}\par\nobreak\medskip\nobreak}
%
%       \eqn\label{a+b=c}	gives a displayed equation with number
%				chosen consecutively within sections.
%     \eqnn and \eqna define labels in advance
%
\def\eqnn#1{\xdef #1{(\secsym\the\meqno)}%
\global\advance\meqno by1\eqnlabel#1}
\def\eqna#1{\xdef #1##1{\hbox{$(\secsym\the\meqno##1)$}}%
\global\advance\meqno by1\eqnlabel{#1$\{\}$}}
\def\eqn#1#2{\xdef #1{(\secsym\the\meqno)}\global\advance\meqno by1%
$$#2\eqno#1\eqlabel#1$$}
%
%			 footnotes
\newskip\footskip\footskip14pt plus 1pt minus 1pt %sets footnote baselineskip
\def\f@@t{\baselineskip\footskip\bgroup\aftergroup\@foot\let\next}
\setbox\strutbox=\hbox{\vrule height9.5pt depth4.5pt width0pt}
\global\newcount\ftno \global\ftno=0
\def\foot{\global\advance\ftno by1\footnote{$^{\the\ftno}$}}
%
%     \ref\label{text}
% generates a number, assigns it to \label, generates an entry.
% To list the refs on a separate page,  \listrefs
%
\global\newcount\refno \global\refno=1
\newwrite\rfile
\def\ref{[\the\refno]\nref}
\def\nref#1{\xdef#1{[\the\refno]}\ifnum\refno=1\immediate
\openout\rfile=refs.tmp\fi\global\advance\refno by1\chardef\wfile=\rfile
\immediate\write\rfile{\noexpand\item{#1\ }\reflabel{#1}\pctsign}\findarg}
%	horrible hack to sidestep tex \write limitation
\def\findarg#1#{\begingroup\obeylines\newlinechar=`\^^M\pass@rg}
{\obeylines\gdef\pass@rg#1{\writ@line\relax #1^^M\hbox{}^^M}%
\gdef\writ@line#1^^M{\expandafter\toks0\expandafter{\striprel@x #1}%
\edef\next{\the\toks0}\ifx\next\em@rk\let\next=\endgroup\else\ifx\next\empty%
\else\immediate\write\wfile{\the\toks0}\fi\let\next=\writ@line\fi\next\relax}}
\def\striprel@x#1{} \def\em@rk{\hbox{}} {\catcode`\%=12\xdef\pctsign{%}}
\def\semi{;\hfil\break}
\def\addref#1{\immediate\write\rfile{\noexpand\item{}#1}} %now unnecessary
\def\listrefs{\vfill\eject\immediate\closeout\rfile%\parindent=20pt
\baselineskip=12pt\centerline{{\bf References}}\bigskip{\frenchspacing%
%\catcode`\@=11
\escapechar=` \input refs.tmp\vfill\eject}\nonfrenchspacing}
\def\startrefs#1{\immediate\openout\rfile=refs.tmp\refno=#1}
\def\figures{\centerline{{\bf Figure Captions}}\medskip\parindent=40pt}
\def\fig#1#2{\medskip\item{Figure ~#1:  }#2}
\catcode`\@=12 % at signs are no longer letters
%
%	Unpleasantness in calling in abstract and title fonts
\ifx\answ\bigans
\font\titlerm=cmr10 scaled\magstep3 \font\titlerms=cmr7 scaled\magstep3
\font\titlermss=cmr5 scaled\magstep3 \font\titlei=cmmi10 scaled\magstep3
\font\titleis=cmmi7 scaled\magstep3 \font\titleiss=cmmi5 scaled\magstep3
\font\titlesy=cmsy10 scaled\magstep3 \font\titlesys=cmsy7 scaled\magstep3
\font\titlesyss=cmsy5 scaled\magstep3 \font\titleit=cmti10 scaled\magstep3
\else
\font\titlerm=cmr10 scaled\magstep4 \font\titlerms=cmr7 scaled\magstep4
\font\titlermss=cmr5 scaled\magstep4 \font\titlei=cmmi10 scaled\magstep4
\font\titleis=cmmi7 scaled\magstep4 \font\titleiss=cmmi5 scaled\magstep4
\font\titlesy=cmsy10 scaled\magstep4 \font\titlesys=cmsy7 scaled\magstep4
\font\titlesyss=cmsy5 scaled\magstep4 \font\titleit=cmti10 scaled\magstep4
\font\absrm=cmr10 scaled\magstep1 \font\absrms=cmr7 scaled\magstep1
\font\absrmss=cmr5 scaled\magstep1 \font\absi=cmmi10 scaled\magstep1
\font\absis=cmmi7 scaled\magstep1 \font\absiss=cmmi5 scaled\magstep1
\font\abssy=cmsy10 scaled\magstep1 \font\abssys=cmsy7 scaled\magstep1
\font\abssyss=cmsy5 scaled\magstep1 \font\absbf=cmbx10 scaled\magstep1
\skewchar\absi='177 \skewchar\absis='177 \skewchar\absiss='177
\skewchar\abssy='60 \skewchar\abssys='60 \skewchar\abssyss='60
\fi
\skewchar\titlei='177 \skewchar\titleis='177 \skewchar\titleiss='177
\skewchar\titlesy='60 \skewchar\titlesys='60 \skewchar\titlesyss='60
\def\titlefont{\def\rm{\fam0\titlerm}% switch to title font
\textfont0=\titlerm \scriptfont0=\titlerms \scriptscriptfont0=\titlermss
\textfont1=\titlei \scriptfont1=\titleis \scriptscriptfont1=\titleiss
\textfont2=\titlesy \scriptfont2=\titlesys \scriptscriptfont2=\titlesyss
\textfont\itfam=\titleit \def\it{\fam\itfam\titleit} \rm}
\ifx\answ\bigans\def\abstractfont{\tenpoint}\else
\def\abstractfont{\def\rm{\fam0\absrm}% switch to abstract font
\textfont0=\absrm \scriptfont0=\absrms \scriptscriptfont0=\absrmss
\textfont1=\absi \scriptfont1=\absis \scriptscriptfont1=\absiss
\textfont2=\abssy \scriptfont2=\abssys \scriptscriptfont2=\abssyss
\textfont\itfam=\bigit \def\it{\fam\itfam\bigit}
\textfont\bffam=\absbf \def\bf{\fam\bffam\absbf} \rm} \fi
\def\tenpoint{\def\rm{\fam0\tenrm}% switch back to 10-point type
\textfont0=\tenrm \scriptfont0=\sevenrm \scriptscriptfont0=\fiverm
\textfont1=\teni  \scriptfont1=\seveni  \scriptscriptfont1=\fivei
\textfont2=\tensy \scriptfont2=\sevensy \scriptscriptfont2=\fivesy
\textfont\itfam=\tenit \def\it{\fam\itfam\tenit}
\textfont\bffam=\tenbf \def\bf{\fam\bffam\tenbf} \rm}
%
%---------------------------------------------------------------------
%
\def\noblackbox{\overfullrule=0pt}
\hyphenation{anom-aly anom-alies coun-ter-term coun-ter-terms}
\def\inv{^{\raise.15ex\hbox{${\scriptscriptstyle -}$}\kern-.05em 1}}
\def\dup{^{\vphantom{1}}}
\def\Dsl{\,\raise.15ex\hbox{/}\mkern-13.5mu D} %this one can be subscripted
\def\dsl{\raise.15ex\hbox{/}\kern-.57em\partial}
\def\del{\partial}
\def\Psl{\dsl}
\def\tr{{\rm tr}} \def\Tr{{\rm Tr}}
\font\bigit=cmti10 scaled \magstep1
\def\biglie{\hbox{\bigit\$}} %pound sterling
\def\lspace{\ifx\answ\bigans{}\else\qquad\fi}
\def\lbspace{\ifx\answ\bigans{}\else\hskip-.2in\fi} % $$\lbspace...$$
\def\boxeqn#1{\vcenter{\vbox{\hrule\hbox{\vrule\kern3pt\vbox{\kern3pt
	\hbox{${\displaystyle #1}$}\kern3pt}\kern3pt\vrule}\hrule}}}
\def\mbox#1#2{\vcenter{\hrule \hbox{\vrule height#2in
		\kern#1in \vrule} \hrule}}  %e.g. \mbox{.1}{.1}
%	matters of taste
%\def\tilde{\widetilde} \def\bar{\overline} \def\hat{\widehat}
%
% some sample definitions
\def\CAG{{\cal A/\cal G}}   %curly letters
\def\CA{{\cal A}} \def\CC{{\cal C}} \def\CF{{\cal F}} \def\CG{{\cal G}}
\def\CL{{\cal L}} \def\CH{{\cal H}} \def\CI{{\cal I}} \def\CU{{\cal U}}
\def\CB{{\cal B}} \def\CR{{\cal R}} \def\CD{{\cal D}} \def\CT{{\cal T}}
\def\e#1{{\rm e}^{^{\textstyle#1}}}
\def\grad#1{\,\nabla\!_{{#1}}\,}
\def\gradgrad#1#2{\,\nabla\!_{{#1}}\nabla\!_{{#2}}\,}
\def\ph{\varphi}
\def\psibar{\overline\psi}
\def\om#1#2{\omega^{#1}{}_{#2}}
\def\vev#1{\langle #1 \rangle}
\def\lform{\hbox{$\sqcup$}\llap{\hbox{$\sqcap$}}}
\def\darr#1{\raise1.5ex\hbox{$\leftrightarrow$}\mkern-16.5mu #1}
\def\lie{\hbox{\it\$}} %pound sterling
\def\ha{{1\over2}}
\def\half{{\textstyle{1\over2}}} %puts a small half in a displayed eqn
\def\roughly#1{\raise.3ex\hbox{$#1$\kern-.75em\lower1ex\hbox{$\sim$}}}

\font\names=cmbx10 scaled\magstep1

\def\sg{{\sigma_8({\rm gal})}}
\def\sc{{\sigma_8({\rm CDM})}}
\def\bias{{\sigma_8^{-1}({\rm CDM})}}
\baselineskip=20pt
\Title{PUP-TH-93/1393}
{\vbox{\centerline
{$\Pi$ in the Sky ?}
\centerline{
 Microwave Anisotropies from Cosmic Defects}}}
%   \footnote{}{*optional footnote on title}
\font\large=cmr10 scaled\magstep3
\font\names=cmbx10 scaled\magstep1
\centerline{ \bf\names David Coulson$^{1,2}$}
\centerline{\bf\names Pedro Ferreira$^2$}
\centerline{\bf\names Paul Graham$^1$}
\centerline{ and }
\centerline{\bf\names Neil Turok$^1$}

\centerline{$^1$ Joseph Henry Laboratories,  Princeton University, Princeton,
NJ08544.}

\centerline{$^2$ The Blackett Laboratory, Imperial College, London SW7, UK. }
\centerline{\bf Abstract}
\baselineskip=12pt

High resolution maps of the anisotropy of the microwave sky
will yield invaluable clues as to the mechanisms
involved in cosmic structure formation.
One fundamental question they should answer
is whether the fluctuations  were
Gaussian random noise, as predicted in inflationary models,
or
were  nonGaussian as in theories based on symmetry breaking and
 cosmic defects.
In the latter case there is the prospect of
obtaining information regarding
symmetry breaking at high energy scales,
specifically the homotopy classes $\Pi_n$ of the vacuum manifold.
In this paper we report on detailed calculations of the
degree scale anisotropies predicted in the cosmic string,
monopole,
texture and nontopological texture theories of structure formation,
emphasising their distinct character from those predicted
by inflation, and the bright prospects for experimental tests.

\Date{10/93} %replace this line by \draft  for preliminary versions
             %or specify \draftmode at some point
\baselineskip=24pt
\centerline{\bf I. Introduction}

The search for a theory of cosmic structure formation
is one of the most exciting areas of  science today.
Recent  observations
have drastically tightened constraints
on theories,
 but most basic questions remain open.
One of these
is whether the primordial perturbations in the universe
took the form of
a random-phase  superposition of linear waves
(i.e. Gaussian noise),
or whether there was important additional structure.
The two simplest
{\it ab initio}
families of theories are quite distinct in this regard.

The first invokes inflation, a hypothetical
epoch of very rapid expansion
which made the universe very large and smooth, to
stretch  microscopic quantum fluctuations to
very large length scales. Essentially because the
fluctuations must be kept small to preserve the homogeneity
of the universe, the generation mechanism is linear
and Gaussian fluctuations are predicted.

The second set of theories involves the breaking
of symmetries in unified field theories.
According to  these theories, symmetries are  broken
as the hot early universe cooled, forming
a disordered phase (like ice forming in rapidly
freezing water). This phase
would contain defects
such as cosmic strings, monopoles or textures
 \ref\Kib{T.W.B. Kibble, J. Phys. {\bf A9} 1387 (1976).}.
 As the universe
expands and cools, the tangle of defects straightens itself out
on progressively larger scales.
Despite the  different physics involved, the inflation and
defect theories are surprisingly  similar in their predictions,
and  detailed calculations and observations have not yet
convincingly ruled either out.  In this paper we argue  that
this state of affairs should  not last long --
degree scale
microwave anisotropy data available in the near future
could
decide the issue.

The cosmic microwave background radiation provides
a uniquely clean probe for cosmic defects.
As the defects evolve,
they
produce a time-dependent gravitational field  which
both disturbs the homogeneity of the universe and
shifts the energy of the cosmic microwave photons
passing near the defects.
A map of the microwave sky would reveal the pattern of
defects in the universe as seen by their influence on the
microwave photons which reach us.
 For
each kind of defect there are specific nonGaussian signatures --
cosmic strings would produce
linelike discontinuities on the sky
 \ref\ks{N. Kaiser and A. Stebbins, Nature
{\bf 310}, 391-393 (1984).},
and texture would
produce hot and cold spots
 \ref\ts{N. Turok
and D. Spergel, Phys. Rev. Lett. {\bf 64}, 2736 (1990).}.

The simplest defect theories are  highly  predictive,
the perturbations being governed by
one parameter, the symmetry breaking scale $\phi_0$,
of order the grand unification
scale, $10^{16}$ GeV.
The theoretical `payoff'
should one of these theories be verified
would be  high -- maps of the microwave sky could yield
information
about symmetry breaking at very high energies,
specifically the homotopy classes $\Pi_n$ of the vacuum
manifold.

The COBE (Cosmic Background Explorer Satellite)
provided the first evidence of anisotropy
on the microwave sky \ref\Smoot{G.F. Smoot {\it et. al. }, Ap. J. Lett.
{\bf 396} L7 (1992).}.
This gives a clean way
to fix the
overall magnitude of density perturbations
in each theory.
Further information is limited by COBE's low resolution -
the large ($10^o$)  beam blurs any information on smaller scales -
and computations show that its findings are compatible
with essentially {\it all}
of the cosmic defect and inflationary theories
\ref\Ben{
D. P. Bennett and S. H. Rhie, Livermore preprint (1992).},
\ref\PST{U. Pen, D. Spergel and N. Turok,  Phys. Rev. {\bf D}
in the press (1993).}.

Higher resolution measurements offer the prospect of definitely
excluding theories, and
answering  the question of Gaussianity.
The data gathered so far
form a partial but nonetheless intriguing picture.
The MAX experiment  observed two regions of the sky and
found
one to be quiet and the other noisy
-- assuming Gaussian statistics,
discrepant
upper and lower 95 \% confidence bounds were obtained
 \ref\Gun{J. O.
Gundersen {\it et. al.}, Ap. J. Lett. {\bf 413} L1 (1993);
P. Meinhold {\it et. al.}, Ap. J. Lett. {\bf 409} L1 (1993).}.
And the South Pole data of Schuster et. al.
 \ref\Sch{J. Schuster {\it et. al}, Ap. J. Lett {\bf 412} L47
(1993).}
contains a statistically significant nonGaussian `bump'
\ref\PG{P. Graham, N. Turok, P.M. Lubin and
J. Schuster, Ap. J. in the press (1993).}.

Furthermore,
the detected temperature differences  $\delta T/T$
vary across a  narrow band of angular scales.
The PYTHON experiment \ref\Drag{M. Dragovan et. al., Princeton preprint
(1993).} detects fluctuations at a level $\sim 3\times 10^{-5}$
on a scale of $\sim 3^o$, and
MAX
at $4\times 10^{-5}$ on a scale of $\sim .5^o$. But
the SP91 experiment of Gaier et. al. \ref\Gai{T. Gaier {\it et al.}, Ap. J.
Lett. {\bf 398} L1 (1992).}
set an upper limit of $1.4\times 10^{-5}$ on a scale of $\sim 1^o$,
and the
Schuster et. al. scan yields an even stronger limit\PG.
These differing results do not appear compatible with
a Gaussian sky with a smooth power spectrum.
It is clearly of interest to determine the predictions
of the nonGaussian theories on these angular scales,
while the observational situation is still being resolved.

Techniques for realistic calculations of  microwave  anisotropies
in the cosmic defect theories
have been recently developed  \Ben, \PST. The idea is to
follow the ordering of the defect fields  numerically,
solve the linearised Einstein equations to
determine the gravitational fields produced, and
simultaneously compute the energy shift for
photons on paths converging at an
`observer' at the current epoch.  Previously, such calculations
have been performed only for large angular scales,
for comparison with COBE. On smaller angular scales,
it is important to include
the `Doppler' effect produced as the photons scatter off
moving electrons.
This effect is likely to be more
Gaussian than the direct gravitational effect of the defects, and
an important question is whether it  produces a noisy enough
background
to mask the distinctive `signatures' of the cosmic
defects. For a realistic assessment of the observability of
those signatures, it is important to include all the effects relevant
on a given angular scale.

A central issue in these calculations is whether the universe was
reionised at high redshifts.
 According to the standard theory,
electrons and protons combined to make neutral hydrogen at a redshift
of $\sim 1000$.
But
observations of quasars show that the intergalactic medium was
almost fully ionised
at a redshift of $\sim 5$,
probably   by the release of ultraviolet radiation
from stars or even black holes.

The main effect of the reionisation for the microwave anisotropy is
to suppress the anisotropy on smaller angular scales. Consider
the photons now arriving at us along a given line of sight.
If we followed each photon back in time along its path, we would
would see it follow a random walk, scattering
increasingly frequently at higher redshift as the electron density grew.
At very early times
the photon position  would be effectively
`frozen' as its mean free path tended to zero.
The photons we see along a given direction
thus probe the gravitational potential in the early universe
over a length scale of order the photon mean free path
at the time when typical photons
`last scatter', of order the
the distance light has travelled since the big bang (the horizon
scale) at that time.
Reionisation, by causing the photons to
scatter at later epochs, smooths the observed pattern of microwave
anosotropy on the angular scale subtended by the horizon
at `last scatter'.
This scale is
of order $2^o$ without reionisation and $6^o$ with reionisation.
(Throughout this paper we assume the `canonical values'
$\Omega=1$, $\Lambda=0$, $h=0.5$, $\Omega_B=0.05$ -
although since for many other values the universe was close
to critical density at the relevant epochs, one can simply
rescale the angular scale of our maps to obtain the
corresponding results.)

Reionisation significantly smooths the microwave
anisotropy only if it occurs sufficiently early.
At redshifts later than $\sim 50$ the density of electrons
is so low that scattering has little effect.
In cosmic defect theories,
reionisation at redshifts higher than $100$ is
quite likely \PST, unlike the case in most
 inflationary  scenarios
 (although see \ref\ST{J. Silk
and M.Tegmark, CfPA 93-th-04 (to appear in Ap.J.).} , \ref\FK{M. Fukugita and
M.
 Kawasaki, ICRR-Report-301-93-13.} ).

\centerline{\bf Techniques: Field and Defect Evolution}

For global monopoles, texture and nontopological texture
accurate  and well tested methods are available to
evolve the fields \PST.
Cosmic strings are more difficult.
After nearly a decade of work uncertainties remain \ref\CSbook{
{\it The Formation and Evolution of Cosmic Strings}, Ed. G. Gibbons,
S. Hawking and T. Vachaspati, Cambridge University Press, 1990.},
stemming from the sensitivity to
small scale structure on the long strings -- different
ways of handling this numerically lead to different scaling
densities
for the long strings.
Existing codes incorporating cosmic expansion
have limited  dynamic range (an expansion factor of less than 20).
We have attempted instead to model string network evolution by making
use of an {\it exact} lattice solution to the Nambu equations
for string evolution and reconnection
first used by Smith and Vilenkin \ref\SV{G. Smith and A. Vilenkin,
Phys. Rev. {\bf D 36} 990 (1987); M. Sakellariadou and A. Vilenkin,
Phys. Rev. {\bf D 42} 349 (1990). }. Mapping coordinates to preserve the
causal structure (time to conformal time, space to comoving coordinate)
one obtains a simple model of evolution in an expanding universe.
This model
\ref\FT{P. Ferreira and N. Turok, in preparation (1993).} exhibits
similar scaling densities and  `fractal' structure to
that of the more complex expanding universe codes, but has the virtue
that
very large simulations  are possible, extending over an expansion factor
in excess of $10^4$,
allowing computations
with large dynamic range.

Prior to this work, Bouchet et. al.
\ref\BBS{F. Bouchet, D. P. Bennett and A. Stebbins,
Nature {\bf 335} (1988) 410.}
constructed a picture of the  anisotropy produced by
cosmic strings by stacking expanding universe simulations
end to end, and using a dipole approximation derived in
Minkowski space
to model  the anisotropy pattern.
This exaggerates the linelike
discontinuities produced by the strings.
We instead use our model string network
as a source for the full linearised Einstein equations,
which we solve to compute the photon energy shifts.
We also include the effects of electron scattering.
Our  conclusion is
that resolution higher than a degree is needed
to see clearly the discontinuities from strings.

\centerline{\bf Photon Scattering}

The first part of our technique is to construct the photon paths
backwards in time, propagating away from the observer.
During a time interval $[t_i,t_f]$, the probability for
a photon not to scatter is given by
\eqn\ps{\eqalign{
&P = e^{-d} \qquad \qquad d= \int_{t_i}^{t_f} dt n \sigma_T = n_o \sigma_T t_o
[ ({t_o\over t_i}) - ({t_o\over t_f})]
}}
where $ n_e$ is the electron density,  $\sigma_T$ the
Thompson cross section,  $t$ cosmic time and the subscript
 $o$ refers to the present
\ref\Pee{P.J.E. Peebles, Ap. J. {\bf 315}
 L73 (1987).}.
We apply \ps\ at each timestep of
the photon path, drawing a random number to determine whether
a scattering actually occurred. If it did, a new direction
is chosen at random for the photon, and the scheme iterates.
When the photon mean free path falls below
our resolution scale, we store
the entire set of photon paths, and begin the main calculation.

We evolve the fields  from uncorrelated initial conditions
with the horizon set equal to one grid spacing. Simultaneously
we solve
the linearised Einstein equations using the analytic
formulae of \PST, obtaining the  perturbation to the
spacetime metric $h_{ij}(\eta, {\bf x})$. This
is then used to compute the  energy shift
the photons experience,  through the `Sachs-Wolfe' formula
\eqn\dl{\eqalign{
&\qquad \qquad {\delta T \over T}({\bf n}) =
-{1\over 2} \int_i^f d\eta h_{ij,0}\bigl(\eta, {\bf x}(\eta)\bigr) n^i n^j}}
where ${\bf x}(\eta) $ is the photon path, and $\eta$ the conformal time
coordinate.

We include the
`Doppler' term mentioned above in the  approximation
that at the epochs of interest the electron velocity is equal
to the velocity of the dark matter.
This  is reasonable because photon drag is small
at the redshifts of interest
 \ref\Pee{P.J.E.
Peebles, {\it Principles of Physical Cosmology}, Princeton University
Press, 1993.} .
We have extended the analytic treatment of \PST, to show
how the `Doppler' term emerges from \dl, and may be
calculated with similar precision to the time dependent
potential term\ref\CFGT{D. Coulson, P. Ferreira, P. Graham and
N. Turok, in preparation (1993). }.

A typical run employs a box of $128^3$ for the field evolution
($320^3$ for strings),
$64^3$ for the metric perturbations and photon trajectories,
with 100 photons along each of $64^2$ lines of sight.
Six sets of photons are used, each set converging from
a square subtending a $30^o$ square  to  an `observer'
at the centre of
one of the faces of the simulation volume.
The code has been extensively tested to check that
it reproduces known analytic
solutions for anisotropies caused by straight strings, oscillating
loops, and collapsing textures. We have
performed exhaustive tests for finite size and resolution
effects, which will be reported elsewhere \CFGT.
Finite grid effects
are certainly present, but become small
on scales corresponding to a few grid spacings and above.
We  smooth our final maps
on a one degree scale (four field grid spacings, eight string grid spacings)
and believe they are trustworthy above these scales.

\centerline{\bf Results}

Sky maps of the CMB anisotropy produced by cosmic defects
are shown in
Figure 2 a-d. Figure 2e shows an equivalent map
for the simplest inflationary model, commonly referred to  as
standard CDM (cold dark matter) theory. This theory
provides us with a useful Gaussian benchmark.
All maps are $30^o$ on a side and have been smoothed
with a Gaussian window of FWHM $1^o$.

In Figure 1, we show the power  spectra of spherical harmonics on
the sky, computed by averaging over
eighteen maps obtained from 3 independent simulations for each theory.
For comparison
the power spectra for standard CDM,
with and without early reionisation are also shown.
All theories are normalised to fit COBE (as corrected in
\ref\Wright{E. Wright et. al. COBE preprint 1993.}) on large angular scales,
assuming $\Omega=1$, $\Lambda=0$.
At small $l$ the defect theory results are consistent with previous
all sky  simulations \PST. The error bars
show the
`cosmic variance' -- clearly many such patches
on the sky will be necessary to obtain a good estimate of
the power spectrum at small $l$.
there would be more power at smaller angular scales.

There is a striking suppression of power at higher $l$,
most significant in the texture and
nontopological texture theories. This is partly due to reionisation
but also because
 the rms velocities induced in the defect
theories are relatively small compared to
those in inflation.
Inflationary fluctuations
are `built in'  whereas in the defect theories they are `induced'
by motion of the defect fields, requiring
larger gravitational fields for a given velocity
\PST.

In Table 1 we show a quantity more closely related to
what experiments  measure, the
rms temperature gradients in each theory.
We caution that these are {\it strongly dependent}
on the values of as yet poorly determined cosmological parameters,
$h$, $\Lambda$ and $\Omega$.
In an open universe, or
in a $\Lambda $-dominated flat universe, there would be some
suppression
of the anisotropy on COBE scales,
with a relative increase in  the power on degree scales.
Likewise, if reionisation were less than 100\% efficient,
there would be more power on smaller scales.

\centerline{\bf NonGaussianity}

The main focus of our work has been to determine whether
the nonGaussianity of the defect theories makes them
readily distinguishable from Gaussian scenarios.
The maps provide visible indications of this -
the texture and monopole maps  have peaks and troughs in excess of
five standard deviations, with a characteristic structure.
For textures some maps like Figure 2c
show clear, isolated hot and cold spots a few degrees across
as expected \ts.
For monopoles,
all of our maps show many pairs of hot regions separated by
a colder region, a pattern one expects from
an annihilating monopole-antimonopole pair. Most of
these are small, but a few are still visible even after
degree-scale smoothing.
Figure 2b shows one large monopole-antimonopole
pair annihilating to the left and right of the cold spot
left of centre of the picture, and one smaller diagonal pair just below
that.
The  string maps do show evidence of lines along which the temperature
changes sharply from hot to cold, but they are
not overwhelmingly  different from standard CDM.

Objective statistical criteria are needed
to determine whether a given sky
map is nonGaussian or not.  As our basic method,
we take each sky map and randomise the phases of the
Fourier components, thus constructing an ensemble of
1000
Gaussian maps with the same  power spectrum.
We then search for statistics  which strongly distinguish the original
 maps from the Gaussian ensemble. We found
measures based on the distributions of the temperature itself  to be
rather weak.
The simplest
are
the the skewness $< T^3>/<T^2>^{3\over 2}$ and kurtosis
$<T^4>/<T^2>^2-3$, both
zero for a Gaussian distribution. Using eighteen  maps
for each theory, we calculated the mean skewness and
kurtosis of the temperature.
None were statistically significant.
Likewise the maximum temperature was not  a  very powerful
discriminator, because a significant number of the maps {\it do not}
have extreme  peaks.

In previous work \PG, we suggested that {\it large gradients} might
be a better way  to
distinguish the nonGaussian from Gaussian theories, and we have found this to
be a powerful idea in the present study. One simple test is
as follows. A map is considered `significantly nonGaussian'
if
the maximum gradient in it is  larger than that in
$99\%$ of the ensemble of Gaussian maps based upon it.
Table 2 shows the fraction (out of eighteen) of the
maps satisfying this criterion for each theory.
Most of the texture and monopole maps were strongly
nonGaussian
by this criterion, the string maps much less so.

One would prefer
a statistic
less sensitive to
extreme data points, but the
the significance of the nonGaussianity increases with
increasing resolution, and it is
the small scale structure of the
relatively rare
defect induced signatures that is clearly the most significant
feature.
This
suggests that the optimal
experimental strategy to establish nonGaussianity would  be to measure
temperature gradients at lower (one degree) resolution,
and then `focus in'
on regions of high gradient with higher resolution.

For strings,
the evidence for nonGaussianity is weak.
The sharp discontinuities produced by strings
are not readily apparent  on a one degree smoothing scale.
On smaller scales, a realistic calculation
would require  techniques which
include the effects of Doppler scattering more accurately.

\centerline{\bf Conclusions}

Our conclusion is that the goal of distinguishing
between the texture and monopole  theories
from Gaussian theories
based on inflation should be achievable from microwave
anisotropy
data available soon.
One   difference is in the spectrum
of spherical harmonics, due both to reionisation and
the difference between `induced' and `built-in'
cosmological perturbations. This is very dependent
on other cosmological parameters however.
More distinctive
is the nonGaussianity of the anisotropy pattern
- this  may even be detectable by
degree scale experiments currently in progress.

\centerline{\bf Acknowledgements}
We thank U. Pen for collaboration in
the early stages of this project, and particularly for his
help with computational aspects.
DC thanks the SERC (UK) for financial
support, while PF thanks Programa Ciencia (Portugal).
The work of NT was partially  supported by the SERC (UK),
NSF contract PHY90-21984,   and
the David and Lucile Packard Foundation.

\listrefs

\baselineskip=12pt

\magnification=\magstep1
\def\doublespace{\baselineskip=20pt}
\def\singlespace{\baselineskip=10pt}
\def\lbaselines{\baselineskip=10.10pt   %definition of line spacing
                     \lineskip=0pt
                     \lineskiplimit=0pt}
\def\oneskip{\vskip\baselineskip}         %abbreviation for skipping one line
\def\blankline{\oneskip}
\hoffset=1.92 truecm    %horizontal offset (in addition to the standard 1 inch)
\voffset=1.2  truecm    %vertical offset (in addition to the standard 1 inch)
\hsize=15 truecm      %horizontal pagesize
\vsize=22 truecm      %vertical pagesize
\nopagenumbers

\baselineskip=16pt

\centerline{\bf Table 1. Degree scale $(\delta T/T)'s$}
$$
\vbox{
\halign{# \hfil & \quad # \hfil
&  \hfil # & \quad \hfil #  \cr
\noalign{\hrule}
\noalign{\smallskip}
\noalign{\hrule}
\noalign{\medskip}
{Theory} & {$(\delta T/T)_{1.5^o}$ } & { $(\delta T/T)_{3^o}$  } \cr
\noalign{\medskip}
\noalign{\hrule}
\noalign{\medskip}
\cr
Strings & $2.4 \pm 0.5 \times 10^{-5}$ & $ 2.4\pm 0.5 \times 10^{-5 }$\cr
Monopoles & $1.6 \pm 0.3 \times 10^{-5}$ &$ 1.9 \pm 0.4\times 10^{-5}$\cr
Textures &$ 0.9 \pm 0.2\times 10^{-5}$& $ 1.4 \pm 0.3 \times 10^{-5}$\cr
N.T. Textures & $ 0.9 \pm 0.2\times 10^{-5}$ & $1.4 \pm 0.3 \times 10^{-5}$\cr
Standard CDM &$2.4 \pm 0.5\times 10^{-5}$ & $1.9 \pm 0.4 \times 10^{-5}$\cr
Reionised CDM&$1.4\pm 0.3\times 10^{-5}$ & $1.7 \pm 0.3 \times 10^{-5}$\cr
\cr
\noalign{\hrule}
\noalign{\smallskip}
\noalign{\hrule}
\noalign{\smallskip}}}$$
$$\vbox{
\noindent
Table 1. The typical values of the map gradients
on degree scales in each theory considered here.
The sky maps are first smoothed with
a Gaussian of FWHM $\theta$. The standard deviation of
the two dimensional
 temperature {\it gradient}
$\sigma~\equiv~T^{-1}~\langle~
|~\vec{\nabla}T~|~^2~\rangle^{1\over 2}$
 is then calculated.
The table shows $\sigma \theta$,  the
typical
temperature gradient on the scale $\theta$,
for $\theta$  equal to $1.5$ and  $3$
degrees.
 All theories
have been normalised to fit the standard deviation on
the 10$^o$ scale reported by COBE, $(\delta T/T)_{10^o}= 1.24 \pm
0.2 \times 10^{-5}$.
This yields the values
$G\mu = 2.0\times 10^{-6}$ for strings, and $8\pi G\phi_0^2  = 6.7 \times
10^{-5}$, $1.1 \times 10^{-4}$ and $1.8\times 10^{-4}$ for
monopoles, textures and nontopological textures respectively.
Note that {\it  these numbers are  for the `canonical' parameters
$h=0.5$, $\Omega=1$, $\Lambda=0$, and fractional reionisation $f=1$.}
 }
$$
\baselineskip=16pt

\centerline{\bf Table 2. NonGaussianity in Cosmic Defect Anisotropy Maps}
$$
\vbox{
\halign{
# \hfil & \quad
# \hfil & \quad
# \hfil
&  \hfil # & \quad \hfil # \cr
\noalign{\hrule}
\noalign{\smallskip}
\noalign{\hrule}
\noalign{\medskip}
{Theory} & {$0^o$} & {$1.2^o$}  & { $2.4^o$} & {$3.6^o$} \cr
\noalign{\medskip}
\noalign{\hrule}
\noalign{\medskip}
\cr
Strings & 4& 2 & 2 & 0\cr
Monopoles & 18 & 18 & 18 & 13\cr
Textures &18&  17 & 6 & 2\cr
N.T. Textures &4 & 1 & 1 & 1\cr
Standard CDM &  0&0 & 0& 0\cr
\cr
\noalign{\hrule}
\noalign{\smallskip}
\noalign{\hrule}
\noalign{\smallskip}}}$$
$$\vbox{
\noindent
Table 2. The number of maps out of eighteen constructed for
each theory that were `significantly nonGaussian',
defined as follows. Maps passing the test were those which
had a larger maximum temperature gradient (after smoothing
with a Gaussian of FHWM 1.2, 2.4 or 3.6$^o$) than each of
99\% of the
Gaussian ensemble of maps created by randomising the phases of the
Fourier components of each map.
A similar procedure could in principle be followed with observational
data.
 }
$$

\baselineskip=12pt \vfill
\centerline{\bf Figure Captions}

\fig{1}{ Anisotropy power spectra for the cosmic
defect theories.
The temperature pattern on each map is
expanded as a sum of spherical harmonics, $(\delta T/T)(\theta,\phi) =
\Sigma a_{lm} Y_{lm}(\theta ,\phi )$, with $\theta$, $\phi$
the usual polar angles. The power spectrum
$c_l$ is defined as the ensemble average $<|a_{lm}|^2>$. The power
per logarithmic interval in $l$ is then $l^2 c_l$, constant
for a scale invariant pattern.  All theories considered here
are approximately scale invariant for harmonics with
$l < 10$, to which COBE is sensitive. Differences appear at higher $l$.
The coloured curves show the average power spectrum for
eighteen cosmic defect maps, with the error bars showing the
standard deviation (the `cosmic variance'). The solid curve
shows the spectrum for the `standard CDM' model, and
the dashed line that for the same theory with early reionisation
assumed.
}

\fig{2}{ Temperature anisotropy maps in the
string, monopole, texture, nontopological texture
and standard CDM
scenarios.  The side of the box subtends an angular scale
of $30^o$, and
$\delta T/T$ is given in units of standard deviation for each
map.
All  maps
have been smoothed with a Gaussian of FWHM $1.0^o$.
The nongaussianity is quite apparent in the texture and monopole
maps (note the five sigma peaks and troughs), much less so
in the string and nontopological texture theories.
}

\bye